\documentclass[conference]{IEEEtran}
\usepackage{cite}

\ifCLASSINFOpdf
   \usepackage[pdftex]{graphicx}
   \graphicspath{{../pdf/}{../jpeg/}}
\else
\fi
\usepackage{amsmath}
\usepackage{array}

\begin{document}
%
\title{Network biomarkers of schizophrenia by graph theoretical investigations of Brain Functional Networks}



\author{\IEEEauthorblockN{Megha Singh}
\IEEEauthorblockA{Department of Electrical Engineering\\
Indian Institute of \\ Technology Jodhpur\\
Email: pg201384008@iitj.ac.in}
\and
\IEEEauthorblockN{Rahul Badhwar}
\IEEEauthorblockA{Department of Biology\\
Indian Institute of \\ Technology Jodhpur\\
Email: pg201384010@iitj.ac.in}
\and
\IEEEauthorblockN{Ganesh Bagler}
\IEEEauthorblockA{Center for Computational Biology,\\Indraprastha Institute of \\ Information Technology Delhi,\\
DA-IICT Gandhinagar,\\
Indian Institute of Technology Jodhpur\\
Email: bagler@iiitd.ac.in}}

\maketitle

\begin{abstract}
Brain Functional Networks (BFNs), graph theoretical models of brain activity data, provide a systems perspective of complex functional connectivity within the brain. Neurological disorders are known to have basis in abnormal functional activities, that could be captured in terms of network markers. Schizophrenia is a pathological condition characterized with altered brain functional state. We created weighted and binary BFN models of schizophrenia patients as well as healthy subjects starting from fMRI data in an effort to search for network biomarkers of the disease. We investigated 45 topological features of BFNs and their higher order combinations (2 to 8). We find that network features embodying modularity, betweenness, assortativity and edge density emerge as key markers of schizophrenia. Also, features derived from weighted BFNs were observed to be more effective in disease classification as compared to those from binary BFNs. These topological markers may provide insights into mechanisms of functional activity underlying disease phenotype and could further be used for designing algorithms for clinical diagnosis of schizophrenia as well as its early detection.
\end{abstract}



%
\IEEEpeerreviewmaketitle

\section{Introduction}
Neuronal disorders are known to have basis in abnormal brain functional activities. Brain imaging data have been used to investigate underlying structure and function of a healthy brain and also to pin down differences in functional activity under pathological conditions such as schizophrenia, Alzheimers and autism. Beyond identification of neuronal correlates of these disorders, the need to identify patterns in functional activity has paved way for systems modeling of brain activity data and search for higher order features.

Amongst several neuro-imaging techniques, fMRI has gained widespread popularity for scrutinizing brain activity, owing to its high spatial and temporal resolutions. The fMRI data is collected in terms of voxels where each voxel corresponds to hemodynamic response of the neural activity. This \mbox{four-dimensional} \mbox{spatio-temporal} fMRI data could be used to create systems-level models of brain activities using graph theoretic (complex networks) approach~\cite{Bullmore2011,DeVicoFallani2012,Fox2010}.

Brain functional networks (BFNs) are graph theoretical models of functional activities that provide a deep visual insight into connectivity patterns within the brain. BFNs could be utilized to measure anatomical or functional connectivity between different brain regions and hence to probe network characteristics of functional connectivity under brain disorders with the help of graph theoretical metrics. There is a growing interest in application of BFN models for studying various cognitive states as well as pathological conditions and development of methods for the same. In last decade, several advances have happened towards application of network theory for investigation of fMRI data~\cite{Bullmore2009}. This approach has been used to understand the organization of brain at macro-level using BFN models~\cite{Fair2007,Fair2009,Power2010}. A variety of network modelling approaches have been implemented for this purpose~\cite{Smith2011,Kaiser2011}. These studies have provided insights into the systems architecture of the brain and have highlighted salient features such as presence of default mode network in resting state of brain~\cite{Greicius2003}, small-world architecture~\cite{Achard2006,Bassett2006}, modularity and hierarchical organization~\cite{Zhou2006,Meunier2009,Meunier2010}.

Significant structural and functional neuronal abnormalities are known to happen under pathological conditions such as schizophrenia~\cite{Liu2008,Bluhm2007,Yu2011a}, autism~\cite{Kennedy2008,Monk2009}, and Alzheimer’s~\cite{Stam2007a}. There is an increased focus in finding potential network biomarkers for brain disorders. This will not only assist clinical diagnosis but could eventually help in early diagnosis and effective treatment at reduced cost. 

Schizophrenia, known for altered functional brain state, has been studied with the help of BFN models and machine learning techniques for its classification from healthy brain states. Anderson and Cohen classified schizophrenia patients from healthy under resting and tasked activities using distance matrices modeled from ICAs of fMRI scans with classification accuracies up to 90$\%$~\cite{Anderson2013}. Yang \emph{et al.} demonstrated hybrid machine learning method using fMRI and genetic data of schizophrenic patients for their classification with high accuracies~\cite{Yang2010}. Similar studies were performed using features extracted from default mode network and motor temporal ICA components employing two level feature detection technique~\cite{Du2012}. Using linear and non-linear discriminative methods. Resting state functional connectivity features were examined for disease classification to achieve up to 96$\%$ accuracy using non-linear classifiers~\cite{Arbabshirani2013}. Multiple kernel learning was further used by \cite{Castro2014} to modify the feature selection method thereby achieving  improved accuracies. Chyzhyk \emph{et al.} used extreme learning machines to build a computer aided diagnostic system employing features derived from fMRI data~\cite{Chyzhyk2015}. Beyond these key studies, many efforts have gone into application of machine learning techniques on fMRI derived brain network parameters for classification of schizophrenia with higher accuracy~\cite{Castro2011,Hayasaka2013,Silva2014,Pouyan2014,Savio2015,Du2015}.

In this study we investigated BFNs constructed from fMRI data of schizophrenia subjects and healthy subjects so as to identify higher order topological features that characterize the disease. Starting with dataset provided by center for biomedical research excellence (COBRE), we constructed BFN models for their graph theoretical investigations. Towards identification of key distinguishing network features of schizophrenia, we exhaustively investigated 17 and 28 first order derived network features of binary (unweighted) and weighted BFNs, as well as their higher order tuples. We believe that features thus identified could be effectively used for semi-automated diagnosis of schizophrenia, and may further be used for early detection protocol.

\section{Materials and Methods}
For investigating systems-level differences in brain activity of healthy subjects and pre-diagnosed schizophrenic patients we used the COBRE dataset. This dataset, obtained from International Neuroimaging Data-Sharing Initiative under 1000 Functional Connectomes Project, comprised of fMRI data of 148 subjects. Out of which 74 were healthy subjects (controls) and 72 were patients prediagnosed with schizophrenia and 2 subjects (0040070 and 0040083) were marked as disenrolled, with age groups ranging from 18 to 65 years in all classes. In initial steps of preprocessing, it was found that 2 subjects (0040075 and 0040126) had discrepancies in their data (time points 67-149 were not available and the data was of high frequency nature respectively) which led their removal from the subjects under study. With this, there were 70 patients and 74 healthy controls. The patients were pre-diagnosed with schizophrenia based on `structured clinical interview' used for DSM disorders. Echo-planar imaging was used for resting state fMRI data collection with (Repetition Time) TR=2s, (Echo Time) TE=29ms, matrix size:~64$\times$64, slices=32, voxel size=3$\times$3$\times$4 mm$^3$. 


The fMRI data is a 4-Dimensional metadata comprising of information from spatial and temporal dimensions of blood oxygen-level dependent activities under resting state of the brain. We implemented a protocol to preprocess fMRI data as described in 1000 functional connectome project~\cite{Biswal2010}. To convert preprocessed raw fMRI data we decomposed 4-D data into spatial and time series components using ICA and creating functional graphs from the components by forming corelation matrix~\cite{Anderson2013}. Finally we evaluated graph theoretical properties of brain functional networks analysied it in combinations to find its role as diagnostic biomarkers for schizophrenia.

\subsection{Preprocessing}

\begin{figure}[!t]
\centering
\includegraphics[scale=0.25]{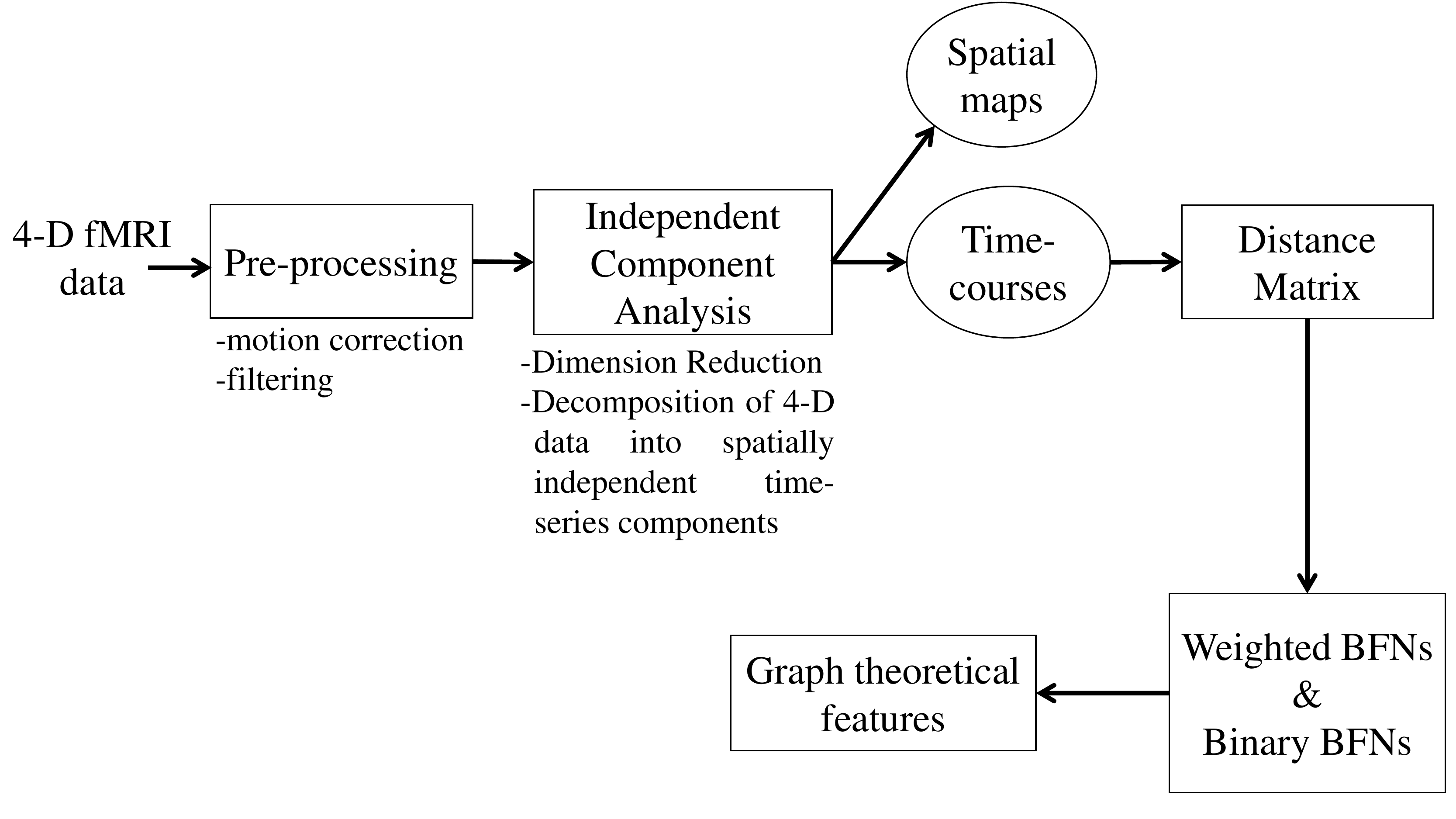}
\caption{Strategy implemented for modeling 4-D fMRI data as brain functional network and its topological characterization.}
\label{fig_flowchart}
\end{figure}

For construction of functional networks the COBRE data needs to be processed before further analysis. The preprocessing step is aimed at removal of artifacts and standardizing locations of brain regions across all subjects. As a first step of pre-processing, we performed deobliquing such that anatomical and time series data correspond to each other. The data is then oriented to match it to a reference master dataset making xyz grid coordinates similar in all the subjects. Motion correction was implemented for each subject by synchronizing signals recorded in each voxel across all slices. This step corrects any misrepresentation of data from individual voxels due to subject's head motion. The corrected recording from each voxel was obtained by shifting images from each slice with respect to the reference image. To eliminate extra tissue around the brain skull stripping was performed, this is done by expanding spherical surface to envelop the brain and exclude rest of the tisues. This step was then followed by smoothing and filtering steps, which include spatial smoothing and temporal filtering to remove high frequencies enhancing low frequencies. To reduce the noisy signals we used grand-mean scaling for normalising the data and applying band-pass filtering. The lowpass and highpass cut-off frequency was set to 0.1 Hz and 0.005 Hz respectively using fast fourier transform of the data. To focus on fluctuations in the signal, removal of linear and quadratic trends was carried out. To identify potential area of interest masking of the preprocessed data was carried out. This brain data had to be registered in accordance with standard brain for which we used linear registration process. Segmenatation was performed over the data to acquire information about grey matter, white matter and cerebro spinal fluid in brain. Finally we applied nuisance signal regression to remove nuisance signal and noises thus obtaining filtered 4-D fMRI data for each subject. Pre-processing steps were performed using packages from Analysis of Functional NeuroImages (AFNI) and fMRI Software Library (FSL)~\cite{Cox1996, Smith2004}.

\begin{figure*}[!th]
\centering
\includegraphics[scale=0.75]{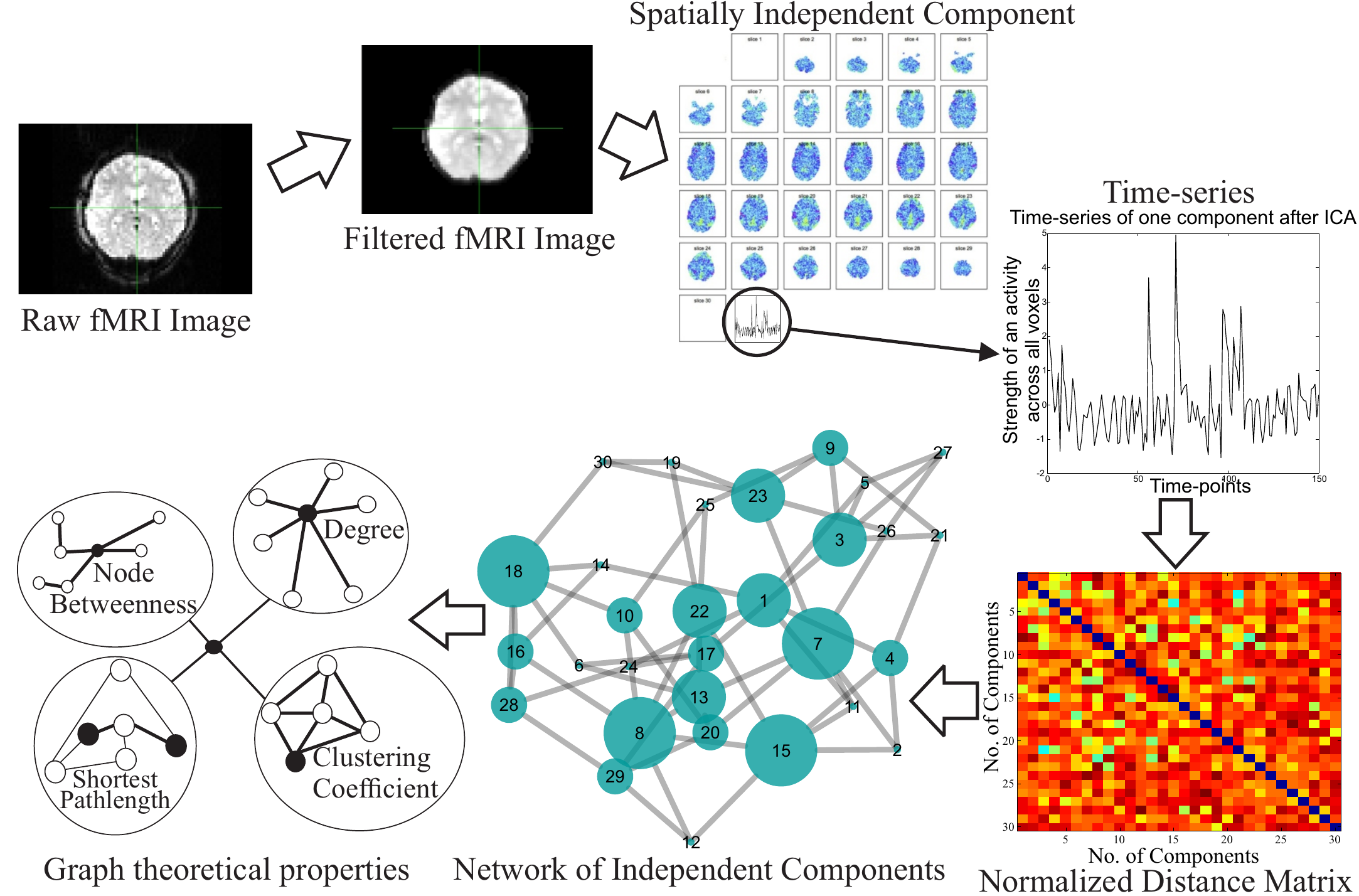}
\caption{Detailed procedure implemented for pre-processing of raw fMRI data, creation of brain functional network of independent components and its characterization using graph theoretical metrics.}
\label{fig_graphflow}
\end{figure*}

\subsection{Independent Component Analysis (ICA)}
Extracting meaningful features from this high dimensional fMRI data is expected to reduce the redundancy and noise that is not removed at the pre-processing stage. Towards this end ICA was implemented to bring down the complexity of this high dimensional data to a manageable level~\cite{Biswal1995,McKeown2003}. The four dimensional space-time fMRI data was represented in terms of an array of dimension $(T,~x*y*z)$ such that the fMRI scan of time length $T$ and space $S$ can be represented by a linear combination of $M<T$ components and corresponding time series: $X_{ts}=\sum_{\mu = 1}^{M} A_{t \mu}C_{\mu s}$ where $X_{ts}$ represents raw scan intensity at time $t$ and space point $s$, $A_{t\mu}$ is the amplitude of component $\mu$ at time $t$, $C_{\mu s}$ is the magnitude of component $\mu$ at space point $s$ and $M$ stands for total number of components. For the COBRE data the time points $T$ in the signal were 150 and $x$, $y$ and $z$ were $64$, $64$ and $32$ where $x$ and $y$ are number of points in two dimensional space and $z$ is the slice number.\\
In ICA, data is assumed to be a linear combination of signals and fMRI data complies with this assumption. Spatial ICA was employed to decompose fMRI data into a set of maximally spatially independent maps and their corresponding time-courses. These time-courses show considerable amount of time-dependencies between distinct functional activities captured in various components, indicating their potential for use in functional network connectivity analysis. 
The time-courses, that measure time-varying activity of components, were used for further analysis. These components represent spatially independent \mbox{time-varying} functional activities of the brain under resting state. 
The ICA of fMRI data was implemented using FSL.

\subsection{Brain Functional Networks and enumeration of network parameters}

Brain functional network is a complex networks model of brain where a node represents `spatially' independent functional activity’ extracted with ICA and an edge specifies the `extent' of temporal dependency between the nodes’. Temporal dependencies between functional activities were measured by finding correlations between them. Dependencies were computed using a correlation based distance metric that is a transformation of the maximal absolute cross-correlation between two time-series. Correlations were calculated for each pair of nodes using cross-correlation function (CCF) over a range of temporal lags,

\begin{equation*}
CCF(X_{\mu_i},X_{\mu_j},l))= \frac{E[(x_{\mu_i,t+l} - \overline{X_{\mu_i}})(x_{\mu_j,t} - \overline{X_{\mu_j}})]}{\sqrt{E[(x_{\mu_i,t} - \overline{X_{\mu_i}})^2]E[(x_{\mu_j,t} - \overline{X_{\mu_j}})^2]}}
\end{equation*}
\\
where, $X_{\mu_i}$, $X_{\mu_j}$ are time series of $i^{th}$ and $j^{th}$  components, $l$ is the temporal lag between them and varied from 0 to 3 points (total 6 seconds with an interval of 2 seconds).  The distance matrix dist($X_{\mu_i}$, $X_{\mu_j}$) was determined by subtracting maximal absolute CCF from 1 and is given by,

\begin{equation*}
d(X_{\mu_i},X_{\mu_j}) = 1 - max[\vert CCF(X_{\mu_i},X_{\mu_j},l) \vert]
\end{equation*}

This distance signifies temporal similarity between two components; the higher the distance lesser the correlation. The distance matrix, thus calculated, represents the weighted brain functional network of a subject. 


The weighted BFN was further pruned to remove weak connections on the basis of k-nearest neighbor approach where k was chosen as 10$\%$ of total components of the subject or 2, whichever was maximum. This pruned BFN was used for computation of graph theoretical parameters. While the graph creation was implemented by $R$ programming language~\cite{TheRCoreTeam2012}, computation of graph theoretical metric was done in MATLAB. \figurename~\ref{fig_flowchart} shows the strategy implemented to obtain BFN of each subject and for its graph theoretical characterization.

Binary BFNs were obtained by thresholding the weighted BFNs. The threshold was chosen on trial and error basis such that a single component network is maintained. Following network parameters were computed on both the binary as well as weighted BFNs for each subject: degree, edge density, node betweenness, edge betweenness, clustering coefficient, characteristic pathlength, efficiency, modularity, closeness, coreness, eccentricity and weak ties. This study was aimed at identification of network parameters that play crucial role in discriminating BFNs of schizophrenia subjects from those of controls. Towards this end we computed higher order statistics (mean, median and standard deviation) for the above mentioned network parameters. We obtained 28 and 17 such statistical features for weighted and binary BFNs, respectively. Table~S1 in Supplementary Material provides an exhaustive list of all network derived features.

\subsection{Feature Classification}

The BFNs were classified into schizophrenic and control subjects on the basis of the network features derived from weighted as well as binary BFNs. Support Vector Machine (SVM) was used as a classifier with radial basis function as its kernel function. When trained with features of BFNs along with their predefined classes, SVM can classify test cases of BFNs. The performance of SVM classification was assessed with 10-fold cross validation statistics. 

We investigated the feature set consisting of individual network features (of weighted and binary BFNs) as well as their 2-8 and 2-6 combinations with a total of 144 subjects. Combinations with more than 8 and 6 networks features were not only computationally challenging ($^{n}C_{2}$ n is the number of features and C is the order of feature combination) but were also found to be redundant towards identification of optimal feature set in binary and weighted BFNs respectively. To identify topological features of BFNs with potential for accurate classification between schizophrenia and healthy subjects by an exhaustive search of network features and their higher combinations we did not exclude any feature. Investigation of each feature's contribution towards classification was carried out by enumerating the features appearance in best 100 accuracies within each feature combinations (2-8 in binary BFNs and 2-6 weighted BFNs).

\section{Results}

\subsection{Brain Functional Networks}

Brain Functional Networks represent systems model of brain functional activities.
\figurename~\ref{fig_graphflow} depicts detailed process used for creation of BFNs starting from raw fMRI data. The raw fMRI data was pre-processed to obtain a filtered fMRI signal which was further decomposed into spatially independent components to fetch time-series for each component using ICA. Using time-series data in components, correlation based normalized distance matrix was calculated. This matrix was transformed into weighted and binary adjacency matrices, which represent the network of independent components. These networks have $28.10(\pm2.21)$ average number of nodes and $47.33(\pm11.72)$ average number of edges between them, thus making graph theoratical properties comparable. Various graph theoretical properties of BFNs were further computed so as to generate a unique feature set to be used for classification (Table~S1 of Supplementary Material). These final features set comprised of combinations of one to eight features derived from binary and upto combinations of one to six features for weighted BFNs. The classification between healthy and schizophrenic subjects was performed with the help of these feature sets.

\begin{figure}[!b]
\centering
\includegraphics[scale=0.45]{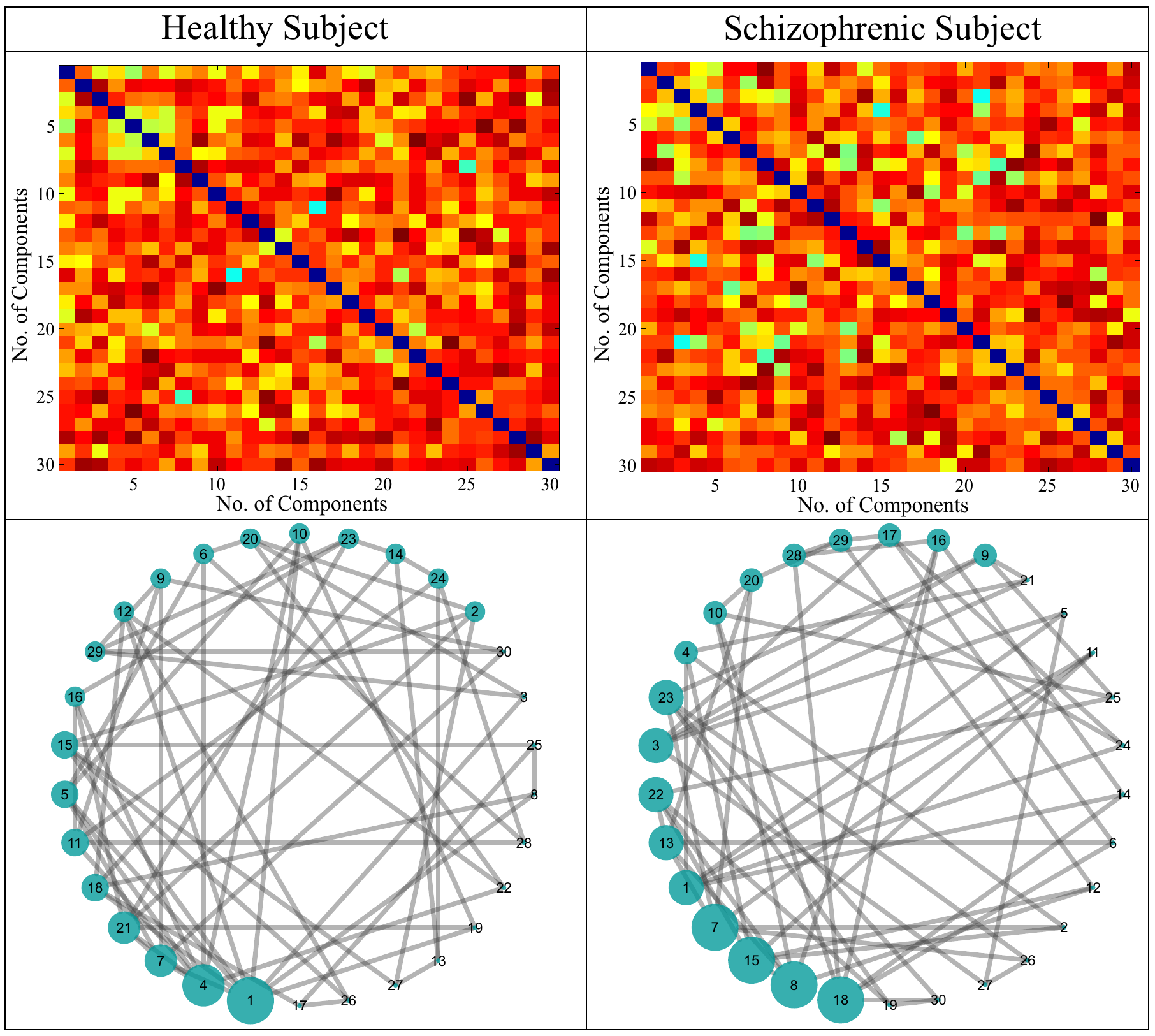}
\caption{Normalized distance matrices and their corresponding weighted BFNs for a representative healthy control (40018) and schizophrenic patient (40009). In the distance matrices lighter gray colors represent lower distances thus higher correlation. The weighted BFNs with 30 components (nodes) show nodes sizes scaled to `clustering coefficient'. The larger the node, higher is its clustering coefficient. The networks were visualized using Cytoscape 3.2.0~\cite{Shannon2003}. 
}
\label{mixed_networks}
\end{figure}

\begin{figure*}[!t]
\centering
\includegraphics[scale=0.35]{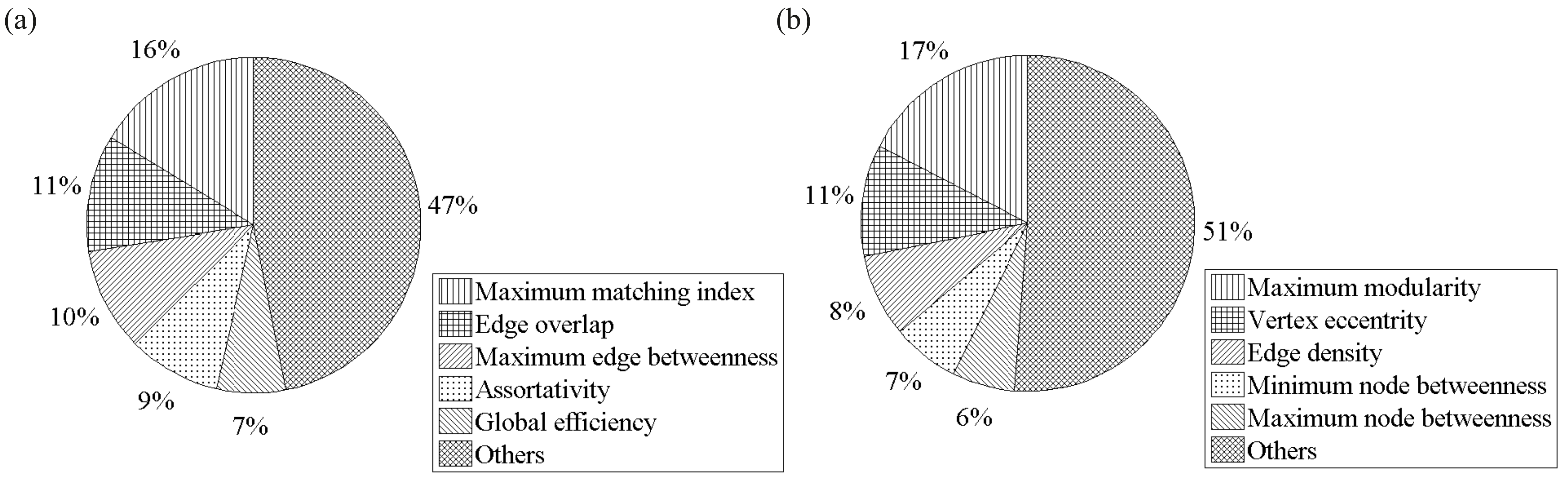}
\caption{Percentage contribution of features with combined discrimination accuracy above 50$\%$ in binary and weighted BFNs (Table~S1 of Supplementary Material). `Others' represents remaining parameters in both types of BFNs.}
\label{binary_weighted}
\end{figure*}

\subsection{Topological biomarkers of schizophrenia}
Towards identification of topological biomarkers of schizophrenia, we investigated network features derived from BFNs of patients and healthy controls. 
\figurename~\ref{mixed_networks} illustrate differences between healthy and schizophrenic subjects with the help of distance matrices and weighted BFNs created in synchronization with a topological property (clustering coefficient). The differences that are almost indiscernible at the level of distance matrices, become more apparent when seen through the lens of topological features. 

First, a feature set using individual parameters was created. For weighted and binary BFNs 28 and 17 such parameters were independently trained and tested in the classifier. Motivated with the idea of creating a better feature set that could potentially enhance the ability to distinguish between correlations across the BFNs, we created combinations ranging from 2 to 6 parameters for weighted and 2 to 8 parameters for binary BFNs. Interestingly, the classification performance for higher order feature combination was much better than that of individual features. We increased the order of combination until the accuracies were declining or the number of combinations become exponantially large. To identify the contribution of each feature in combinations, top 100 accuracies were fetched from each set and their individual contributions were calculated over all combinations.

\subsection{Top contributing features in binary and weighted BFNs}
Among the 17 binary and 28 weighted BFNs features, the features whose combined contribution was more than 50$\%$ in classifying healthy and schizophrenic subjects are shown in ~\figurename~\ref{binary_weighted} for (a) binary and (b) weighted BFNs. Maximum matching index and maximum modularity were among the few topological features that contributed significantly to classification accuracy in binary and weighted networks respectively. Matching index quantifies similarity between two nodes' connectivity profiles, excluding their mutual connection. Modularity quantifies the degree to which the network may be subdivided into delineated groups. The difference in networks belonging to two categories support the hypothesis that recognizes schizophrenia as a disorder of dysfunctional integration between distant brain regions~\cite{Bullmore1997,Friston2005}. While the ability to consistently classify schizophrenia BFNs from that of healthy subject may be limited with single features, we anticipated better efficacy for higher order ($>2$) features.

\subsection{Efficacy of feature combinations}

\begin{figure}[!b]
\centering
\includegraphics[scale=0.25]{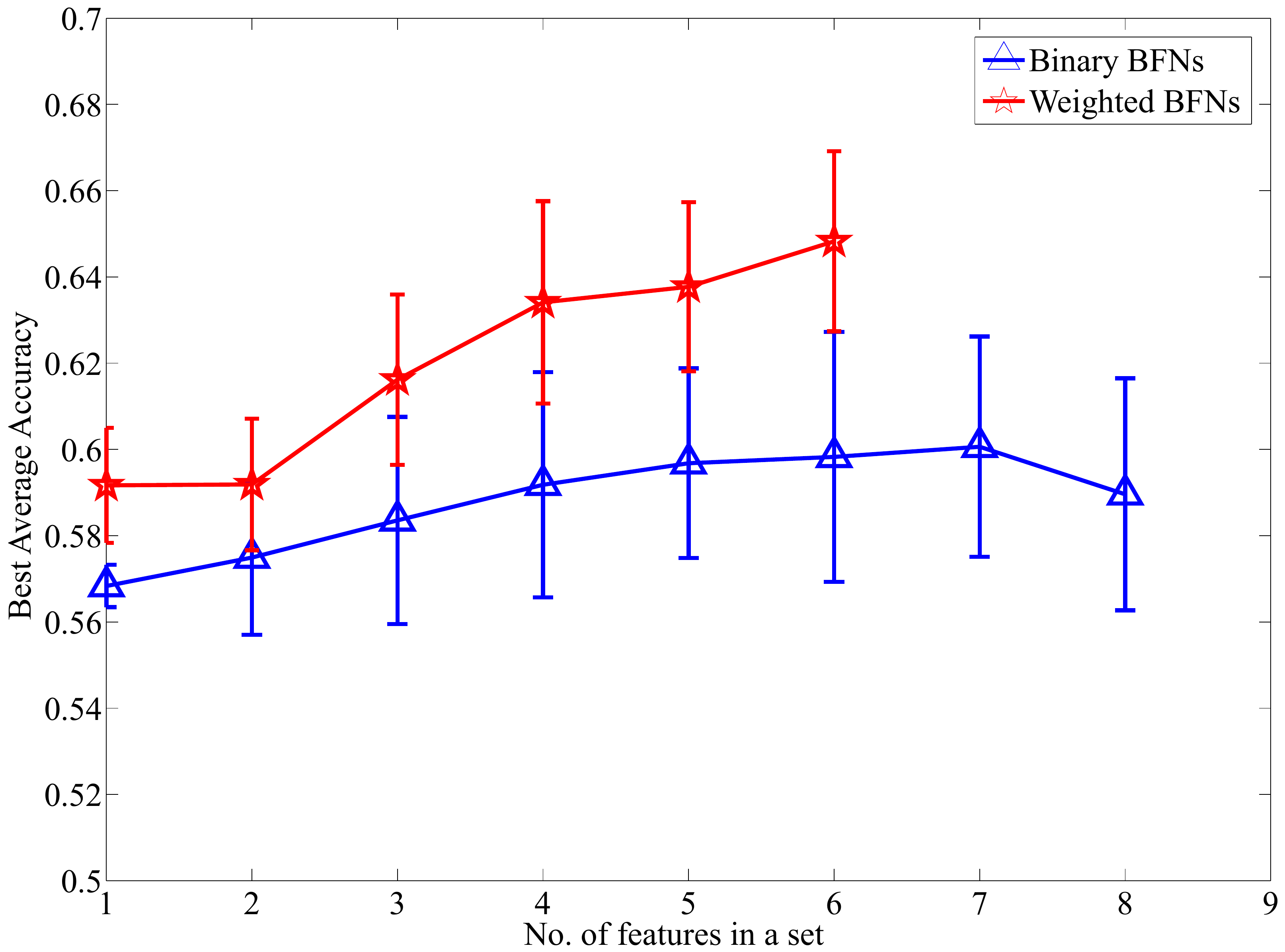}
\caption{Best accuracy obtained for classification between BFN models of healthy and schizophrenia brain data with increased order of feature combinations. The error bars represent standard error for $100$ experiments of $10$-cross fold tests.}
\label{fig_accuracy_trend}
\end{figure}

It was observed that increasing the order of feature combinations yields improvement over individual or pairwise sets. This trend continues till combination of 8 and 6 for binary and weighted BFNs respectively.

\figurename~\ref{fig_accuracy_trend} summarizes the effectiveness of higher order combinations of network features. The figure shows the trend in highest average accuracy with increasing number of features in the feature set. Improved classification accuracy was observed with increased order of features upto 8 combinations in binary BFNs and 6 combinations in weighted BFNs. Interestingly, features extracted from weighted BFNs yielded better classification accuracy than those from binary BFNs. With increased complexity of features the accuracy, after initial increase, plateaued.  Increased redundancy was observed among the top features obtained.

\section{Discussion}
Brain functional networks, graph theoretical models of brain activity data, provide macro-level understanding of complex functional connectivity in the brain~\cite{Bullmore2009,Rubinov2010,Bullmore2011,DeVicoFallani2012}. Functional networks of brains ink pathological conditions, such as schizophrenia, have been reported to have altered properties quantifiable in terms of topological features (For eg.: low average clustering, long characteristic pathlength, lower degree of connectivity, lower strength of connectivity and  reduced modularity)~\cite{Liu2008,Bassett2008,Alexander-Bloch2010,Lynall2010,Yu2011}. Such disruptions of topological features are understood to be an indication of dysfunctionality in schizophrenic BFNs as studied in fMRI scans of dorsal and ventral prefrontal, anterior cingulate, and posterior cortical regions~\cite{Ellison-Wright2008,Rish2013}. 

Here, we performed an exhaustive investigation of graph theoretical features of binary and weighted BFNs and their higher order combinations towards classification of BFNs of schizophrenia and healthy subjects. One of the objectives of our study was to assess the utility of increased order of features on classification accuracy. Other than that we also aimed to identify network metrics and their possible implication for altered brain functional patterns. 

Our study provides some of the key features that can play an important role in characterizing schizophrenia. Liu \emph{et al.} had shown that small-world organization in brain networks of schizophrenic patients are significantly altered in many brain regions with decreased clustering and increased characteristic path length~\cite{Liu2008}. Beyond disrupted small world nature, our study presents other properties that may be associated with BFNs of dysfunctional schizophrenic phenotype. Weighted BFNs of schizophrenia subjects were distinct in terms of maximized modularity, sum of product of degrees across all edges, vertex eccentricity and betweenness. On the other hand, binary BFNs presented matching index, edge overlap, edge betweenness, assortativity and global efficiency among the best features that could be used for classification of schizophrenia patients from healthy subjects. Broadly these properties reflect on connectivity, modularity and hierarchical organization of the network.

One of the highest accuracies reported with BFN-based models equals 65$\%$ as reported by Anderson and Cohen~\cite{Anderson2013}. Although, the classification accuracies achieved in our study were not exemplary, our study highlights the role of topological features derived from weighted BFNs for classification of schizophrenia fMRI data. The study also underlines limits on higher order feature combinations indicating saturation of classification accuracy. Specific network features obtained from our study could further be used for designing better disease classification algorithms as well as early detection systems. 

Our study suggests that instead of exhaustive search for features through higher order combinations of features (n-tuples) appropriate use of feature selection methods could be the way forward. The insights gained from our study into network biomarkers and limitations of higher order features could be used for efficient design of computational protocols for diagnose of schizophrenia at an early stage.

\vfill

\newpage
\onecolumn

\section*{SUPPLEMENTARY MATERIAL}

\setcounter{table}{0}
\renewcommand{\thetable}{S\arabic{table}}

\begin{table}[!ht]
\centering
\caption{Network based features of each subject that were used for classification between two classes.}
\begin{tabular}{|l|l|l|l|}
\hline
S.No.                & Weighted Network Property                  & S.No. & Binary Network Property  \\
\hline
1.                   & Node count                                 & 1.    & Edge count                                  \\
2.                   & Minimum degree                             & 2.    & Edge density                                \\
3.                   & Maximum degree                             & 3.    & Assortativity                               \\
4.                   & Mean degree                                & 4.    & Mean node betweenness                       \\
5.                   & Median degree                              & 5.    & Maximum node betweenness                    \\
6.                   & Standard deviation of degrees              & 6.    & Standard deviation of node betweenness      \\
7.                   & Edge density with threshold                & 7.    & Maximum edge betweenness                    \\
8.                   & Assortativity                              & 8.    & Longest distance between two vertices       \\
9.                   & Mean node betweenness                      & 9.    & Global efficiency                            \\
10.                  & Maximum node betweenness                   & 10.   & Average clustering                        coefficient  \\
11.                  & Minimum node betweenness                   & 11.   & Transitivity                                  \\
12.                  & Standard deviation of node betweenness     & 12.   & Maximum edge overlap                          \\
13.                  & Maximum edge betweenness                   & 13.   & Maximum matching index between two vertices   \\
14.                  & Shortest distance between two vertices     & 14.   & Mean closeness                                \\
15.                  & Global efficiency                          & 15.   & Characteristic pathlength                     \\
16.                  & Average clustering coefficient             & 16.   & Number of weak ties                           \\
17.                  & Transitivity                               & 17.   & Minimum node betweenness                      \\
18.                  & Optimal community structure                &       &                                               \\
19.                  & Maximized modularity                       &       &                                               \\
20.                  & Maximum of matching index between two vertices    &     &                 \\
21.                  & Median coreness                            &       &                                    \\
22.                  & Mean closeness                             &       &                                               \\
23.                  & Median closeness                           &       &                                               \\
24.                  & Standard deviation of closeness            &       &                                               \\
25.                  & Sum of product of degrees across all edges &       &                                               \\
26.                  & Mean Vertex eccentricity                   &       &                                               \\
27.                  & Standard deviation of node betweenness     &       &                                               \\
28.                  & Longest distance between two vertices      &       &                                              \\
\hline
\end{tabular}

\label{all_properties}
\end{table}

\section*{Network Properties}
These network properties are calculated using the Brain Connectivity Toolbox in MATLAB, created by Rubinov and Sporns (M. Rubinov and O. Sporns, `Complex network measures of brain connectivity: uses and interpretations', Neuroimage, vol.\ 52, no.\ 3, pp.\ 105969, Sep.\ 2010.). Following is the list of properties provided as part of the toolbox:
\\
1.	Degree\\
Number of edges connected to a node is called degree. Average of degrees of all nodes is called average degree.
\\~\\
2.	Edge density\\
It is the ratio of edges present in the network to the possible number of edges i.e it is the fraction of present connections to possible connections. So, if `k' is the edges present and `n' is the nodes in the network. Then edge density is: $=\dfrac{k}{\dfrac{n(n-1)}{2}}$
\\~\\
3.	Node betweenness\\
Node betweenness centrality is the fraction of all shortest paths in the network that contain a given node. Nodes with high values of betweenness centrality participate in a large number of shortest paths.
\\~\\
4.	Assortativity\\
The assortativity coefficient is a correlation coefficient between the strengths (weighted degrees) of all nodes on two opposite ends of a link. A positive assortativity coefficient indicates that nodes tend to link to other nodes with the same or similar strength.
\\~\\
5.	Edge betweenness\\
Node betweenness centrality is the fraction of all shortest paths in the network that contain a given node. Nodes with high values of betweenness centrality participate in a large number of shortest paths.
\\~\\
6.	Efficiency\\
The global efficiency is the average of inverse shortest path length, and is inversely related to the characteristic path length. The local efficiency is the global efficiency computed on the neighborhood of the node, and is related to the clustering coefficient.
\\~\\
7.	Optimal community structure and Maximized modularity\\
The optimal community structure is a subdivision of the network into non overlapping groups of nodes in a way that maximizes the number of within-group edges, and minimizes the number of between-group edges. The modularity is a statistic that quantifies the degree to which the network may be subdivided into such clearly delineated groups.
\\~\\
8.	Coreness\\
The k-core is the largest subgraph comprising nodes of degree at least k. The coreness of a node is k if the node belongs to the k-core but not to the (k+1)-core. This function computes the coreness of all nodes for a given binary undirected connection matrix.
\\~\\
9.	Closeness\\
It reflects proximity of a node to the core of the network.
\\~\\
10.	Eccentricity\\
It is the maximum distance of a node to any other node.
\\~\\
11.	Edge overlap\\
It measures the degree of connectivity of two nodes and is the relative number of neighbors of two nodes which are neighbors of each other.
\\~\\
12.	Clustering Coefficient\\
The weighted clustering coefficient is the average "intensity" of triangles around a node.
\\~\\
13.	SOP\\
The sum of products of degrees across all edges

\end{document}